\documentclass[acmtog,screen,balance=true]{acmart}
\setcopyright{none}
\renewcommand\footnotetextcopyrightpermission[1]{}

\usepackage{etoolbox}

\makeatletter
\renewcommand{\@journalName}{}
\renewcommand{\@acmVolume}{}
\renewcommand{\@acmNumber}{}
\renewcommand{\@acmArticle}{}
\renewcommand{\@acmYear}{}
\renewcommand{\@acmMonth}{}

\patchcmd{\runningfoot}
  {\def\@runningfoot{\@journalName, Vol.\ \@acmVolume, No.\ \@acmNumber, Article\ \@acmArticleNum, Publication date:\ \@acmMonth\ \@acmYear.}}
  {\def\@runningfoot{}}
  {}{}

\patchcmd{\firstfoot}
  {\def\@firstfoot{\@journalName, Vol.\ \@acmVolume, No.\ \@acmNumber, Article\ \@acmArticleNum, Publication date:\ \@acmMonth\ \@acmYear.}}
  {\def\@firstfoot{}}
  {}{}

\AtBeginDocument{
  \def\@runningfoot{}
  \def\@firstfoot{}
  \fancypagestyle{plain}{%
    \fancyhf{}
    \fancyfoot{}
    
  }
  \fancypagestyle{firstpagestyle}{%
    \fancyhf{}
    \fancyfoot{}
    
  }
  \pagestyle{plain}
}
\makeatother

\usepackage{booktabs} 

\usepackage[ruled]{algorithm2e} 

\SetAlFnt{\small}
\SetAlCapFnt{\small}
\SetAlCapNameFnt{\small}
\SetAlCapHSkip{0pt}

\usepackage{mathtools}
\usepackage{tabularx}
\usepackage[percent]{overpic}
\DeclareUnicodeCharacter{2194}{\ensuremath{\leftrightarrow}}
\usepackage[utf8x]{inputenc}

\usepackage{multirow}
\usepackage{siunitx}
\usepackage{xspace}
\usepackage{pict2e} 
\usepackage{graphicx} 

\definecolor{ourcol}{rgb}{0.37,0.54,0.28}
\definecolor{condition}{HTML}{FBB040}
\newcommand{\methodname}{PRISM\xspace} 

\begin{document}
\title{\methodname: A Unified Framework for \\ \textcolor{purple}{P}hotorealistic \textcolor{orange}{R}econstruction and \textcolor{lime}{I}ntrinsic \textcolor{cyan}{S}cene \textcolor{violet}{M}odeling}

\author{Alara Dirik}
\email{a.dirik22@imperial.ac.uk}
\orcid{0000-0002-4946-1313}
\affiliation{
    \institution{Imperial College London}
    \country{United Kingdom}
}

\author{Tuanfeng Wang}
\email{yangtwan@adobe.com}
\orcid{0000-0002-8180-4988}
\affiliation{
    \institution{Adobe Research}
    \country{United Kingdom}
}

\author{Duygu Ceylan}
\email{ceylan@adobe.com}
\orcid{0000-0002-2307-9052}
\affiliation{
    \institution{Adobe Research}
    \country{United Kingdom}
}

\author{Stefanos Zafeiriou}
\email{s.zafeiriou@imperial.ac.uk}
\orcid{0000-0002-5222-1740}
\affiliation{
    \institution{Imperial College London}
    \country{United Kingdom}
}

\author{Anna Frühstück}
\email{fruehstu@adobe.com}
\orcid{0000-0002-3870-4850}
\affiliation{
    \institution{Adobe Research}
    \country{United Kingdom}
}
\renewcommand{\shortauthors}{Zeng et al.}

\begin{teaserfigure}
    \includegraphics[width=\textwidth]{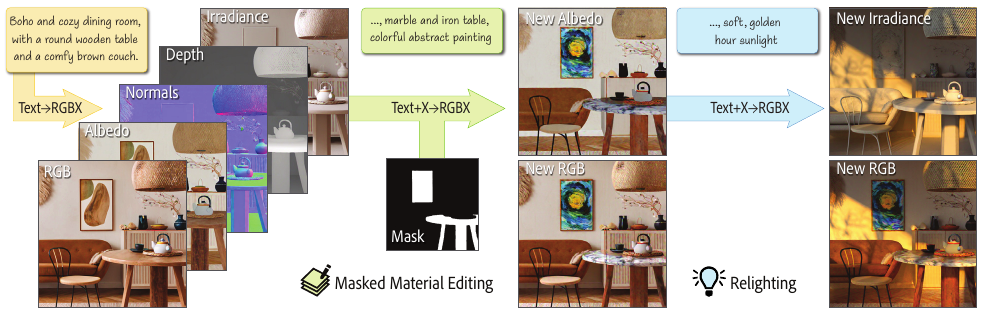}
    \caption{
    We propose \methodname, a unified framework for conditional generation of RGB image and its intrinsic channels (referred to as X layers) simultaneously. It supports a variety of tasks, including text-to-RGBX generation, RGB-to-X decomposition, and X-to-RGBX conditional generation. PRISM achieves plausible results on both local material editing on masked region and global image relighting through conditioning on selected intrinsic layers and text prompts.
    }
    \label{fig:Teaser}
\end{teaserfigure}

\begin{abstract}
We present \methodname, a unified framework that enables multiple image generation and editing tasks in a single foundational model. Starting from a pre-trained text-to-image diffusion model, \methodname proposes an effective fine-tuning strategy to produce RGB images along with intrinsic maps (referred to as X layers) simultaneously. Unlike previous approaches, which infer intrinsic properties individually or require separate models for decomposition and conditional generation, \methodname maintains consistency across modalities by generating all intrinsic layers jointly. It supports diverse tasks, including text-to-RGBX generation, RGB-to-X decomposition, and X-to-RGBX conditional generation. Additionally, \methodname enables both global and local image editing through conditioning on selected intrinsic layers and text prompts. Extensive experiments demonstrate the competitive performance of \methodname both for intrinsic image decomposition and conditional image generation while preserving the base model's text-to-image generation capability.
\end{abstract}

\maketitle

\begin{figure*}[t]
    \centering
    \includegraphics[width=\textwidth]{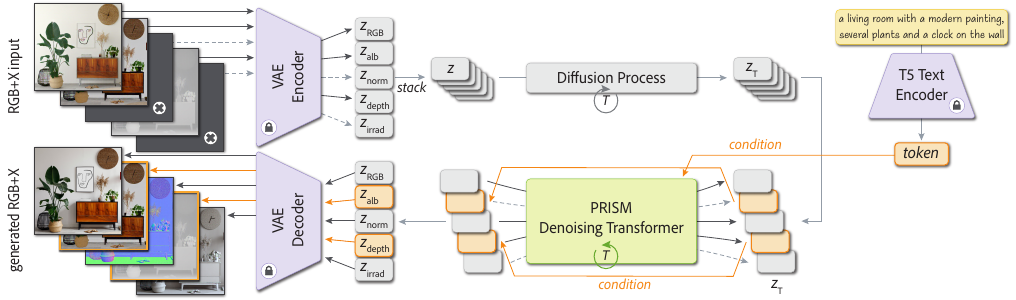}
    \caption{
        \textbf{Pipeline} of our \methodname model. RGB image and its corresponding X intrinsic channels are encoded into latent space via a fixed VAE Encoder. A Diffusion Transformer is applied on the tokens of the latent of all channels simultaneously and conditioned by the text embedding from an input text prompt. Denoised tokens are passing through a fixed decoder for RGB+X generation. During training, intrinsic channels are randomly ablated which makes \methodname a unified framework for text-to-image generation, intrinsic decomposition, and conditional image generation with any subset of intrinsic images.
    }
    \label{fig:pipeline}
\end{figure*}

\section{Introduction}
Foundational image generation models have seen significant advances recently enabling high quality image generation from text descriptions~\cite{ramesh2022hierarchical,Chen2023PixArtFT}. Such models have also been extended for various image editing as well as more explicit conditional generation workflows. Motivated by such rapid success, recent work have proposed to benefit from the generative prior of such models for various perception tasks including depth~\cite{ke2023repurposing} and normal~\cite{ye2024stablenormal} estimation as well as intrinsic image decomposition~\cite{Kocsis2024IntrinsicID,Zeng2024RGBXID,Luo2024}. Such methods often finetune a pre-trained image generation model, effectively replacing the image generation capability with the ability to predict a certain spatial map from a given input image. 

We argue that perceiving an image as a composition of its intrinsic components (albedo, shading, and geometric properties) is a fundamental problem since many image editing tasks (relighting, material editing, and geometric manipulation) can be represented as specific changes in the corresponding intrinsic maps. Hence, instead of replacing the image generation capability with a perception task, we propose that generation and perception tasks should be handled in a unified manner. Such a unified framework preserves the original image generation capability of the foundational model while the generative prior can simultaneously be utilized for perception. 

In this paper, we present \methodname, a unified framework that jointly addresses intrinsic image decomposition and conditional image generation. Starting from a pre-trained diffusion transformer model for image generation, \methodname presents an effective approach to fine-tune the base model to jointly generate RGB images along with its intrinsic maps (which we denote as X layers). Our model enables conditional generation with any combination of X layers while preserving text-based generation. \methodname can perform: (i) text-to-RGBX generation, where text input produces an image with intrinsic layers, (ii) RGB-to-X decomposition from an input image, and (iii) X-to-RGBX generation from a subset of intrinsic layers. Since all X layers are generated jointly, \methodname achieves an improved alignment between the RGB image and each of the intrinsic maps. The third task enables global image editing by conditioning on selected intrinsic maps and text prompts that describe the edits (e.g., for relighting, we condition generation on all the intrinsic maps except the shading layer along with a text prompt that describes the desired lighting). We additionally enable local editing through fine-tuning with masked conditioning maps.

We extensively evaluate \methodname, compare it with strong baselines, and demonstrate competitive performance on multiple tasks. Our method excels in intrinsic image decomposition and consistency across modalities, while also delivering competitive results with respect to visual and conditional generation quality. Even though we use limited datasets when training \methodname, we observe a strong generalization capability when performing RGB-to-X decomposition which we attribute to the unified generation and perception capability. Finally, we show case various downstream applications, such as material editing and relighting, that benefit from our approach. By jointly generating images along with their intrinsic modalities, \methodname facilitates iterative editing of an image while maintaining consistent intrinsic decomposition. In summary, our main contributions include: 1) a unified multi-tasking model for both intrinsic perception and conditional generation that demonstrates strong generalization capability; 2) joint generation of intrinsic layers and RGB images for improved alignment; and 3) competitive performance in intrinsic decomposition and downstream applications.
\vspace{-1em}
\section{Related Work}
\subsection{Text-to-Image Generation}
Text-to-image generation has seen significant advancements in recent years following earlier works on class-conditional and unconditional large-scale image generation models with GAN~\cite{Reed2016GenerativeAT, Brock2018LargeSG, Karras2018ASG, Karras2019stylegan2, Karras2021, Mirza2014ConditionalGA} and diffusion-based~\cite{Ho2020DenoisingDP, Dhariwal2021DiffusionMB} architectures. Earlier works on text-to-image generation predominantly use GANs~\cite{Reed2016GenerativeAT,zhang2017stackgan,xu2018attngan, zhu2019dm, tao2022df, Xia2021TediGANTD}. They either operate in specific domains with limited vocabularies or formulate text-to-image generation as a text-guided editing problem. While works like GigaGAN~\cite{Kang2023ScalingUG} demonstrate that it is possible to scale up GANs as foundational image generators, diffusion models and large-scale language encoders~\cite{Raffel2019ExploringTL, Radford2021LearningTV} have been pre-dominantly used for text-based image generation in the recent years. Works such as GLIDE~\cite{Nichol2021GLIDETP}, DALL$\cdot$E~2~\cite{ramesh2022hierarchical}, and Imagen~\cite{Saharia2022PhotorealisticTD} propose diffusion-based architectures that can generate photorealistic images from text descriptions using a pre-trained frozen text encoder~\cite{Radford2021LearningTV, Raffel2019ExploringTL}. Following the success of U-Net based diffusion models, other variants including transformer-based auto-regressive and diffusion models~\cite{ramesh2021zero,Gafni2022MakeASceneST,ding2021cogview,ding2022cogview2,yu2022scaling,Chen2023PixArtFT} have been widely adopted. We base our work on a pre-trained diffusion transformer-based text-to-image generation model and fine-tune it to jointly generate images along with its intrinsic maps.

\subsection{Image Editing with Diffusion Models}
Several methods have been proposed to add additional conditioning to text-to-image diffusion models to enable custom generation and editing. One approach is to perform large-scale training for task-specific purposes such as domain adaptation and personalization~\cite{Hu2021LoRALA,Ruiz2022DreamBoothFT,Gal2022AnII}. Composer~\cite{Huang2023ComposerCA} formulates image generation as a composition problem and trains a model that takes a text prompt and optional style and content images as input to generate a final image. Kosmos~\cite{Pan2023KosmosGGI, Peng2023Kosmos2GM}, QWEN-VL~\cite{Bai2023QwenVLAF} and OmniGen~\cite{Xiao2024OmniGenUI} propose large scale generative vision-language models that enable image editing via leveraring the model's linguistic reasoning capabilities. Finally, works such as InstructPix2Pix~\cite{Brooks2022InstructPix2PixLT} propose training a diffusion model on input image and edited image pairs with edit instructions to enable instruction-based editing.

In another line of work, methods such as ControlNet~\cite{Zhang2023AddingCC, Zhao2023UniControlNetAC, Li2024ControlNetIC}, T2I-Adapter~\cite{Mou2023T2IAdapterLA},  GLIGEN~\cite{Li2023GLIGENOG} and IP-Adapter~\cite{Ye2023IPAdapterTC} propose training lightweight adapter modules to inject new conditioning information into the latent space of pre-trained diffusion models. They fine-tune the adapter modules while keeping original diffusion models frozen to condition the generation on various modalities such as RGB images, depth maps, skeletal pose images, face identity embeddings~\cite{Papantoniou2024Arc2FaceAF}, etc. 

Finally, works such as Prompt2Prompt~\cite{Hertz2022PrompttoPromptIE}, MasaCtrl~\cite{Cao2023MasaCtrlTM} and DragonDiffusion~\cite{Mou2023DragonDiffusionED} propose attention-based latent manipulation methods to enable fine-grained and targeted edits of both generated and real images. Our unified \methodname framework enables conditioned image generation as well as image editing by iteratively manipulating the intrinsic channels that are either jointly generated or predicted from input images. We provide examples of both global and local editing results in Section~\ref{sec:applications}. 

\vspace{-1em}
\subsection{Intrinsic Image Decomposition}
Intrinsic image decomposition aims to decompose RGB images into appearance and geometry-related channels like albedo, irradiance, depth, and surface normals. Early approaches proposed heuristics that operate on image gradients~\cite{Land1971LightnessAR}, grayscale~\cite{Tappen2005RecoveringII, Tappen2006EstimatingIC,Barron2012ShapeAA}, or RGB images~\cite{Zhao2012ACS,Liu2012StatisticalIF,Liao2013NonparametricFF,Garces2012IntrinsicIB,Gehler2011RecoveringII}, with some leveraging additional input~\cite{Barron2013IntrinsicSP}. These methods relied on optimization based on assumed priors about albedo, illumination, and shape properties.

With the adoption of learning based methods for various graphics and vision tasks, follow up work in intrinsic image decomposition also adopted neural networks to tackle the problem~\cite{narihira2015learning,zhou2015learning, zoran2015learning,fan2018revisiting,liu2020unsupervised, li2018cgintrinsics,jin2022reflectguidance} along with efforts to curate relevant dataset~\cite{bell2014intrinsic,li2018cgintrinsics,li2021openrooms,roberts2021hypersim}. Other recent work~\cite{OrdinalShading, Careaga2024ColorfulDI} focused on illumination and divided the problem into physically-motivated sub-problems that can be modeled with neural networks. In addition to supervised training, methods were proposed that exploit multi-task learning~\cite{kim2016unified,zhou2019glosh,luo2020niid} and unsupervised learning~\cite{li2018learning, lettry2018darn}. 

 Recent work has adapted foundational text-to-image models for inverse tasks like depth~\cite{ke2023repurposing} and normal estimation~\cite{ye2024stablenormal}. In a similar fashion, several methods have been proposed to explore latent directions in pre-trained models like StyleGAN~\cite{Bhattad2023StyleGANKN} or fine-tune diffusion models~\cite{Du2023IntrinsicLA,Kocsis2024IntrinsicID,Luo2024,Zeng2024RGBXID, Zhao2025DICEPTIONAG} to generate intrinsic channels conditioned on an input image. Our work follows a similar spirit, but to the best of our knowledge, \methodname is the first unified model that can simultaneously generate RGB and intrinsic channels while also enabling conditioning with respect to any combination of the channels.
\vspace{-.5em}
\section{Method}
\subsection{Diffusion Models}
\label{sec:diffusion}
Diffusion models~\cite{Ho2020DenoisingDP, Dhariwal2021DiffusionMB, Nichol2021GLIDETP, Peebles2022ScalableDM} are a class of generative models that generate data from Gaussian noise via an iterative denoising process.
Typically, a simple mean-squared diffusion loss is used as the denoising objective:
\begin{equation}
  \mathcal{L}_{\text{simple}} = \mathbb{E}_{\mathbf{x}_0, \mathbf{c}, \epsilon, t}(\|\epsilon - \epsilon_{\theta}(a_t \mathbf{x}_0 + \sigma_t \epsilon, \mathbf{c})\|_2^2),
\end{equation}
where $\mathbf{x}_0$ are training samples with optional conditions $\mathbf{c}$, $t\sim \mathcal{U}(0, 1)$ denotes the diffusion timestep, $~\epsilon\sim \mathcal{N}(0, \mathbf{I})$ is the additive Gaussian noise, $a_t,\sigma_t$ are scalar functions of $t$ determined by the underslying scheduler, and $\epsilon_{\theta}$ is a diffusion model with learnable parameters $\theta$.
\textit{Classifier-free guidance} is most widely employed in recent works~\cite{Nichol2021GLIDETP,ramesh2022hierarchical,Rombach2021HighResolutionIS,Saharia2022PhotorealisticTD} for conditional data sampling from a diffusion model, where the predicted noise is adjusted via:
\begin{equation}
  \hat{\epsilon}_{\theta}(\mathbf{x}_t, \mathbf{c}) = \omega \epsilon_{\theta}(\mathbf{x}_t, \mathbf{c}) + (1 - \omega) \epsilon_{\theta}(\mathbf{x}_t),
\end{equation}

where $\mathbf{x}_t = a_t \mathbf{x}_0 + \sigma_t \epsilon$, and $\omega$ is a guidance weight.
Sampling algorithms such as DDIM~\cite{Song2020DenoisingDI} and DPM-Solver~\cite{Lu2022DPMSolverAF,Lu2022DPMSolverFS} are often adopted to speed up the sampling process of diffusion models.

\subsection{PRISM}
\textbf{Base Model.}
We base our work on a pre-trained diffusion transformer architecture (DiT)~\cite{Peebles2022ScalableDM} that flexibly handles both text and image conditioning through random condition token dropout during training. During training, images are first encoded into the latent space of the diffusion model and then tokenized as patches. Our denoiser architecture provides textual and image embeddings extracted via frozen text and image encoders to the attention layers in each transformer block. Specifically, it follows a standard decoder-only transformer architecture containing self-attention blocks where image and text embeddings are tokenized and appended to the tokens of the image being generated. We follow previous work~\cite{Chen2023PixArtFT, Saharia2022PhotorealisticTD, Esser2024ScalingRF} to employ the T5 large language model~\cite{Raffel2019ExploringTL} as our text encoder. The diffusion timestep is directly added to the image tokens as positional encoding. Once a clean latent image representation is sampled, it is then decoded into the RGB image space. In our experiments, we keep the VAE frozen and only fine-tune the diffusion denoiser. While our base model is flexibly text and image conditioned, \methodname utilizes only the text conditioning, which enables robust performance even in scenarios without text inputs, such as RGB-to-X decomposition tasks.

\textbf{Joint RGBX Generation.} Given the pre-trained text-to-image model, \methodname extends it to generate an RGB image along with its intrinsic properties resulting in a fixed number, $M$, of output images. \methodname enables conditioned generation based on a combination of text prompts and a flexible number of intrinsic maps. We propose an effective fine-tuning strategy without any architectural changes as we discuss next.

Let $T_{base}$ denote the number of input tokens the base model operates on. To accommodate generation with multiple modalities, we expand the input token size to $T_{new}$ as:
\begin{equation}
T_{new} = T_{base} \times M
\end{equation}
where $M$ is the number of modalities ($M = 5$ in our implementation, corresponding to RGB, albedo, normal, depth, and irradiance). Specifically, we encode and patchify each intrinsic map separately and provide a union of all tokens as input to the diffusion network. This expansion enables the tokens of the different modalities to jointly attend to each other in the attention blocks of the diffusion model, and results in output RGB images and intrinsic maps to be spatially aligned. We further adjust the positional encoding of the input tokens such that the network can distinguish which token belongs to which modality. Once a clean set of tokens is generated, we group and decode the tokens of each modality independently. 

We augment \methodname with a flexible conditioning mechanism that enables conditioning on both text and a variable number of intrinsic maps. Let $x_t \in \mathbb{R}^{T_{new} \times d} = \bigcup_{i=1}^M x_t^i$ represent the union of all noisy latent tokens at timestep $t$, where $x^i$ denotes a modality, $M$ is the total number of modalities, and $d$ is the embedding dimension. We denote the set of conditioning modalities as: $\mathcal{C}$ where $0 \leq | \mathcal{C} | \leq M$. At each diffusion step, we perform a slight modification to enable conditional generation. After each noise prediction, we override the tokens of a conditioning modality with its clean counterparts as follows:
\begin{equation}
x_{t}^i = \begin{cases}
x_0^i & \text{if } i \in \mathcal{C} \\
x_t^i & \text{otherwise}
\end{cases}
\end{equation}

These simple changes keep the conditioning modalities unchanged and ensure the remaining modalities are influenced by them. At inference time, \methodname can be utilized for three common tasks: (i) text-to-multi-modality generation by setting $\mathcal{C} = \emptyset$, (ii) intrinsic image decomposition by conditioning on an RGB image, and (iii) conditional image generation by providing at least one of the intrinsic maps as conditions.

\subsection{Implementation Details} 
\label{sec:implementation}
\textbf{Intrinsic channels.} We train \methodname to output the following channels along with RGB images:
\begin{itemize}
\item Surface normal $\mathbf{n} \in \mathbb{R}^{H \times W \times 3}$ specifying per-pixel surface normal defined in the camera coordinates;
\item Depth $\mathbf{d} \in \mathbb{R}^{H \times W \times 3}$ represented as per-pixel disparity in the camera coordinates. We represent the depth as a 3-channel image by replicating the disparity value along the channel dimension.
\item Albedo $\mathbf{a} \in \mathbb{R}^{H \times W \times 3}$, also commonly referred to as base color, which specifies the diffuse albedo for dielectric opaque surfaces; 
\item Diffuse irradiance $\mathbf{e} \in \mathbb{R}^{H \times W \times 3}$, serving as a lighting representation and is defined similar to~\citet{Zeng2024RGBXID}.
\end{itemize}

\textbf{Datasets.} Our approach requires a diverse dataset of paired RGB images and intrinsic channels (normal $n$, depth $d$, albedo $a$, diffuse irradiance $e$), along with text captions for training. Following previous work~\cite{Zeng2024RGBXID}, we combine multiple data sources to fulfill this requirement. We utilize InteriorVerse~\cite{InteriorVerse}, a synthetic indoor scene dataset with over 50,000 rendered images providing $\mathbf{n}$, $\mathbf{d}$, and $\mathbf{a}$ channels. Similar to~\citet{Zeng2024RGBXID}, to mitigate synthetic noise in this dataset, we apply denoising on the rendered images. 
We also use HyperSim~\cite{hypersim}, a photorealistic dataset of more than 70,000 rendered images with $\mathbf{n}$, $\mathbf{d}$, $\mathbf{a}$, and $\mathbf{e}$ channels. To further enhance the realism and diversity of our training data, we use an internal dataset of 50,000 high-quality commercial interior images where only the RGB modality is available. To preserve the text-understanding capabilities of our model, we generate captions for all images in the combined dataset using BLIP-2~\cite{Li2023BLIP2BL}.

\textbf{Training Strategy.} As we train \methodname on datasets with varying modalities available, we adopt several strategies to facilitate training. First, if a modality is not provided in a given dataset, we use zero-padded placeholder images, $x_{t}^m=0$, where $m$ denotes a missing modality. We exclude these missing modalities from the corresponding loss computation. For each batch, we randomly select a subset of available modalities as conditioning signals to effectively perform multi-task training including text-only generation, intrinsic decomposition, and conditional generation.  Finally, we implement data source-consistent batching and stratified sampling to maintain identical percentages of each data source between training and validation sets. This balanced representation of modality combinations and conditioning scenarios improves training stability, given varying modality availability across sources.

\textbf{Inference Settings.} We use \methodname in three main modes including text-to-RGBX generation, RGB-to-X intrinsic decomposition, and X-to-RGBX conditional image generation. For intrinsic decomposition, since we are mostly interested in predicting intrinsic maps of a given RGB image, we find it unnecessary to provide an additional text prompt as input. Hence, in this mode we also disable CFG by setting the guidance weight $\omega=0$. In other modes, we set the guidance weight $\omega=7.5$.

\textbf{Inpainting.} Various image editing tasks focus on locally editing a region, e.g., changing the appearance properties of a particular object as shown in Figure \ref{fig:mat-editing}. To support such cases, we fine-tune \methodname model to support inpainting, i.e., conditioning with masked input maps. Specifically, during training, we simply introduce random masks~\cite{Suvorov2021ResolutionrobustLM} in conditioning maps where we set the latent condition map inside the mask region to be random gaussian noise. When computing the diffusion loss, we discard any masked area in the different maps to enable the network to learn to \emph{inpaint} such regions.
\begin{figure*}[h]
    \centering
    \includegraphics[width=\textwidth]{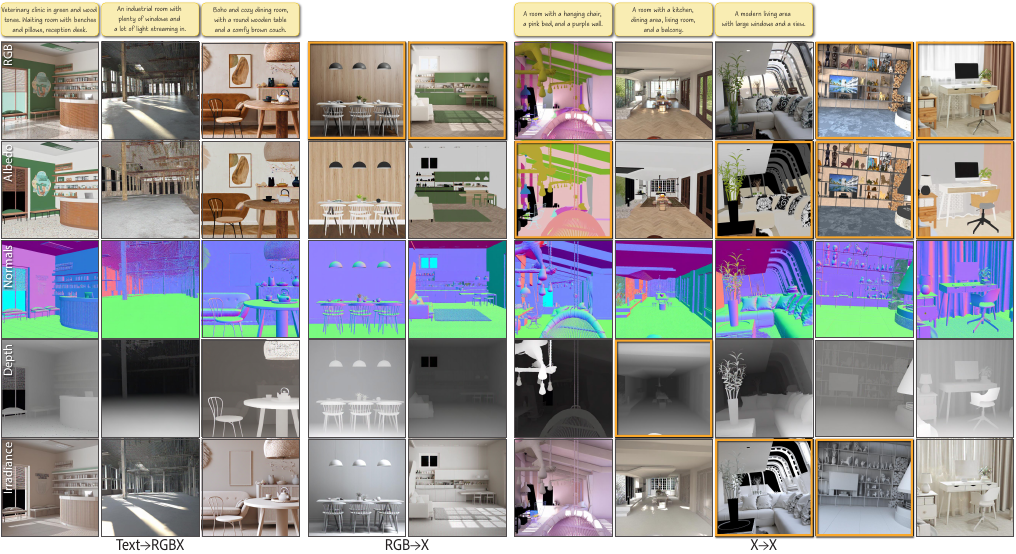}
    \caption{Sample results generated with \methodname. Our model is capable of text, text+X and X conditioned generation and can inherently perform image decomposition and re-composition under different material, geometry and lighting conditions. Condition channels are highlighted in \fcolorbox{condition}{white}{orange}.}
    \label{fig:gen-samples-qual}
\end{figure*}

\section{Experiments}
We perform qualitative and quantitative experiments to showcase the capabilities of \methodname for intrinsic image decomposition, joint image generation, conditional image generation, and editing.

\subsection{Implementation Details}
\textbf{Training and Evaluation Details.}
We fine-tune our base text-to-image model on InteriorVerse \cite{InteriorVerse}, HyperSim \cite{hypersim} datasets, and an internal commercial dataset consisting of real images of interior scenes. We train our model with a batch size of 128 using the AdamW optimizer \cite{Loshchilov2017DecoupledWD} with a weight decay of $0.1$ and a constant \num{5e-5} learning rate, and fine-tune for $15,000$ steps on 8 A100 GPUs. We perform aspect ratio-preserving resizing and random crop of  $512$ × $512$ for training and avoid using a random horizontal flip since it disrupts the camera-space normals. The training of our \methodname model takes around 30 hours and the fine-tuning to enable inpainting takes around 10 hours on 8 A100 GPUs.

For all experiments, we use the official code bases of baseline methods with default settings or use their self-reported metrics where applicable. For conditional image generation experiments, we test our method on the validation set of the pseudo-labeled ImageNet dataset released by \citet{Li2024ControlVAREC}. For consistency with the baseline conditional image generation methods, we downsample our generated results to $256$ x $256$ resolution to compute FID scores. For all conditional image generation baselines, we report their self-reported metrics. For intrinsic image decomposition experiments, we use the official code bases of all baseline methods and run their methods with their respective default settings. For Table \ref{tab:quantitative}, we directly utilize the baseline metrics provided in \citet{Zeng2024RGBXID} on the HyperSim test set and run our method on the same test set. For evaluations on the InteriorVerse dataset, we use our custom train-test split as InteriorVerse does not have a designated test set.


\textbf{Baselines.} We compare our method with state-of-the-art conditional generation and intrinsic image decomposition methods. More specifically, for intrinsic decomposition, we compare against the recent diffusion based ~\cite{Zeng2024RGBXID,Kocsis2024IntrinsicID} and non-diffusion based ~\cite{OrdinalShading, Careaga2024ColorfulDI} methods. We additionally consider an internal method, PVT-normal, based on Pyramid Vision Transformer~\cite{wang2022pvt} and trained on datasets similar to MiDaS~\cite{Ranftl2022, birkl2023midas} to estimate normals. For conditional generation, we compare with ControlNet~\cite{Zhang2023AddingCC}, T2I-Adapter~\cite{Mou2023T2IAdapterLA}, ControlVAR~\cite{Li2024ControlVAREC} and CAR~\cite{Yao2024CARCA}. For depth estimation, we compare our method to a number of state-of-the-art monocular relative and metric depth estimation methods, including Marigold ~\cite{ke2023repurposing} and Depth Anything ~\cite{depth_anything_v1, depth_anything_v2}. Where applicable, we report self-reported metrics of additional baselines for which the codebase or pretrained models are not available.

\textbf{Evaluation Metrics.}
We report Inception Score (IS)~\cite{Salimans2016ImprovedTF} and Frechet Inception Distance (FID)~\cite{Heusel2017GANsTB} to assess the quality of generated RGB images and additionally utilize the CLIP Score~\cite{Hessel2021CLIPScoreAR} to assess the adherence of generated images to the input text prompts. For intrinsic image decomposition, we report the mean PSNR~\cite{zhang2018can} and LPIPS~\cite{Zhang2018TheUE} metrics between the ground truth and generated albedo, normal, and irradiance on the HyperSim test set. To evaluate the depth estimation quality of our method, we report Absolute Relative Error (AbsRel) and Threshold Accuracy ($\delta_1$) metrics on the NYU-v2 \cite{Silberman2012IndoorSA} and ETH3D \cite{Schps2017AMS} datasets. As \methodname is trained to estimate disparity, we first convert our estimated disparity maps to relative depth maps.

\textbf{Qualitative results.} We start by showcasing the capabilities of our model for various tasks in Figure \ref{fig:gen-samples-qual} including text-based joint generation (text-to-RGBX), intrinsic image decomposition (RGB-to-X), conditional generation (X-to-RGBX). As shown, our model can generate realistic and diverse results even in pure text-conditioned cases and adapts to different input condition settings.
\vspace{-.5em}

\subsection{Quantitative Evaluation} \label{sec:quant-eval}
\begin{figure*}[t]
    \centering
    \includegraphics[width=\textwidth]{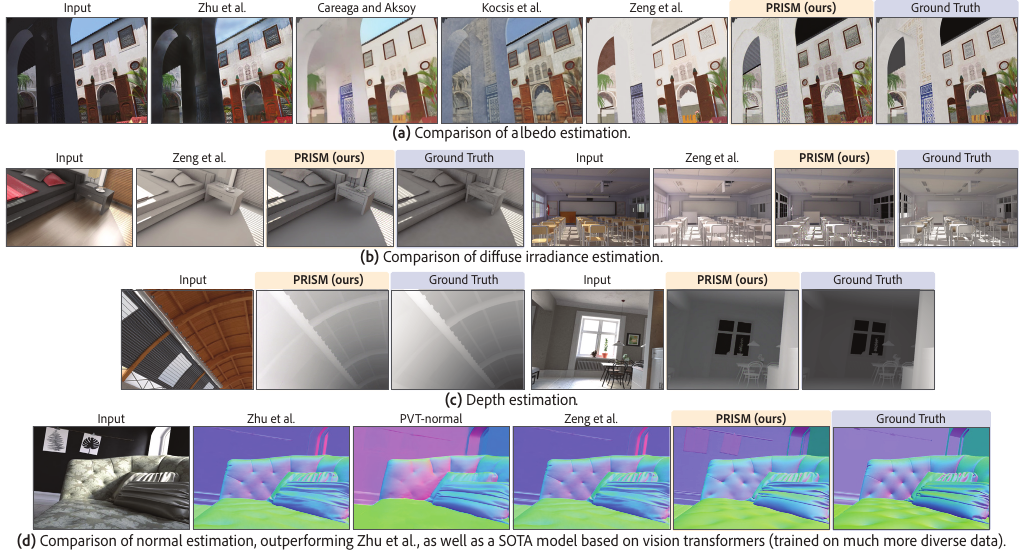}
    \caption{
        Visual comparison of our \methodname model against baseline methods on synthetic datasets. All input images and ground truths are from the HyperSim dataset, except for the classroom scene (b, right).
    }
    \label{fig:rgb2x_synthetic}
    \vspace{-1em}
\end{figure*}

\textbf{Intrinsic image decomposition.}
We show visual intrinsic image decomposition results for synthetic samples in Figure~\ref{fig:rgb2x_synthetic} along with various baseline outputs. For all experiments, we generate $5000$ samples and provide quantitative metrics in Table~\ref{tab:quantitative} where we report the numbers for baseline methods whenever applicable. \methodname performs on-par or superior to state-of-the-art methods while being a unified model capable of multiple tasks.

In order to assess the benefit of jointly estimating multiple modalities on the accuracy of predicting individual maps, we train multiple single image output variants of our method to solely predict albedo, normals, and irradiance given an input RGB image. As shown in Table~\ref{tab:quantitative}, \methodname consistently performs better than its uni-modal variants across all modalities, demonstrating the advantage of joint prediction.

\begin{table}[t]  
    \centering
    \caption{
        Quantitative evaluation of \methodname's intrinsic image decomposition performance against baseline methods on HyperSim test set.
    }
    \label{tab:quantitative}
    \scalebox{0.72}{%
    \centering
        \setlength{\tabcolsep}{2pt}%
        \begin{tabularx}{1.19\linewidth}{ccccccc}
            \toprule
            \multicolumn{1}{c}{\textbf{Method}} & \multicolumn{2}{c}{\textbf{Albedo}} & \multicolumn{2}{c}{\textbf{Normal}} & \multicolumn{2}{c}{\textbf{Irradiance}}                                          \\
            \cmidrule(lr){2-3}
            \cmidrule(lr){4-5}
            \cmidrule(lr){6-7}
            \multicolumn{1}{c}{} & \textbf{PSNR}$\uparrow$ & \multicolumn{1}{c}{\textbf{LPIPS}$\downarrow$} & \textbf{PSNR}$\uparrow$ & \multicolumn{1}{c}{\textbf{LPIPS}$\downarrow$} & \textbf{PSNR}$\uparrow$ & \multicolumn{1}{c}{\textbf{LPIPS}$\downarrow$} 
            \\
            \cmidrule[0.4pt](lr){1-3}
            \cmidrule[0.4pt](lr){4-7}
            \multicolumn{1}{c}{\textbf{\methodname (ours)}} & \textbf{19.3} & \multicolumn{1}{c}{\textbf{0.18}} & \textbf{19.9} & \multicolumn{1}{c}{\textbf{0.18}} & \textbf{18.5} &  \textbf{0.20} \\
            
            \multicolumn{1}{c} {\citet{Zeng2024RGBXID}} & 17.4 & \multicolumn{1}{c}{\textbf{0.18}} & 19.8 & \multicolumn{1}{c}{\textbf{0.18}} & 14.1 & \multicolumn{1}{c}{0.22} \\
            \multicolumn{1}{c}{\citet{InteriorVerse}} & 11.7 & \multicolumn{1}{c}{0.54} & 16.5 & \multicolumn{1}{c}{0.45} & - & -        \\
            \multicolumn{1}{c}{\citet{OrdinalShading}} & 13.5 & \multicolumn{1}{c}{0.34} & - & - & - & \multicolumn{1}{c}{-}                                                   \\
            \multicolumn{1}{c}{\citet{Kocsis2024IntrinsicID}} & 12.1 & \multicolumn{1}{c}{0.41} & - & - & - & \multicolumn{1}{c}{-} \\
            \multicolumn{1}{c}{PVT-normal} & - &\multicolumn{1}{c}{-} & 18.8 & \multicolumn{1}{c}{0.30} & - & - \\
            \multicolumn{1}{c}{\methodname-albedo} & 18.4 & \multicolumn{1}{c}{0.19} & - & -  & - & - \\
            \multicolumn{1}{c}{\methodname-normal} & - &\multicolumn{1}{c}{-} & 18.9 & \multicolumn{1}{c}{0.19} & - & - \\
            \multicolumn{1}{c}{\methodname-irradiance} & - &\multicolumn{1}{c}{-} & - & \multicolumn{1}{c}{-} & 17.9 & 0.21 \\
            \bottomrule
        \end{tabularx}
    }
\end{table}
\vspace{-.5em}

We also evaluate \methodname and the baseline intrinsic decomposition methods on the InteriorVerse test set and report the results in Table~\ref{tab:quant-intverse}. Our method demonstrates superior performance which is consistent to the results shown in Table~\ref{tab:quantitative}.

\begin{table}[h]  
    \centering
    \caption{
        Quantitative evaluation of \methodname's intrinsic image decomposition performance on InteriorVerse test set.
    }
    \label{tab:quant-intverse}
    \scalebox{0.9}{%
    \centering
        \setlength{\tabcolsep}{2pt}%
        \begin{tabularx}{\linewidth}{ccccc}
            \toprule
            \multicolumn{1}{c}{\textbf{Method}} & \multicolumn{2}{c}{\textbf{Albedo}} & \multicolumn{2}{c}{\textbf{Normal}}                \\
            \cmidrule(lr){2-3}
            \cmidrule(lr){4-5}
            \cmidrule(lr){4-5}
            \multicolumn{1}{c}{} & \textbf{PSNR}$\uparrow$ & \multicolumn{1}{c}{\textbf{LPIPS}$\downarrow$} & \textbf{PSNR}$\uparrow$ & \multicolumn{1}{c}{\textbf{LPIPS}$\downarrow$} \\
            \cmidrule[0.4pt](lr){1-3}
            \cmidrule[0.4pt](lr){4-5}
            \multicolumn{1}{c}{\textbf{\methodname (ours)}} & \textbf{19.9} & \multicolumn{1}{c}{\textbf{0.14}} & \textbf{21.2} & \multicolumn{1}{c}{\textbf{0.15}} \\

            \multicolumn{1}{c} {\citet{Zeng2024RGBXID}} & 16.6 & \multicolumn{1}{c}{0.17} & 20.2 & \multicolumn{1}{c}{0.19} \\
            
            \multicolumn{1}{c}{\citet{InteriorVerse}} & 13.6 & \multicolumn{1}{c}{0.24} & 17.1 & \multicolumn{1}{c}{0.26}   \\

            \multicolumn{1}{c}{\citet{OrdinalShading}} & 17.4 & \multicolumn{1}{c}{0.20} & - & -    \\

            \multicolumn{1}{c}{CID} & 17.7 &\multicolumn{1}{c}{0.27} & - & \multicolumn{1}{c}{-}  \\
            
            \multicolumn{1}{c}{\citet{Kocsis2024IntrinsicID}} & 12.2 & \multicolumn{1}{c}{0.30} & 20.1 & 0.21 \\
            
            \multicolumn{1}{c}{\methodname-albedo} & 18.9 & \multicolumn{1}{c}{0.15} & - & - \\

            \multicolumn{1}{c}{\methodname-normal} & - & \multicolumn{1}{c}{-} & 20.4 & 0.16 \\
            
            \multicolumn{1}{c}{PVT-normal} & - &\multicolumn{1}{c}{-} & 17.4 & \multicolumn{1}{c}{0.25}  \\
            \bottomrule
        \end{tabularx}
    }
\end{table}

We separately evaluate the depth estimation quality of our method against a number of state-of-the-art monocular relative and metric depth estimation methods on the NYU-v2 \cite{Silberman2012IndoorSA} and ETH3D \cite{Schps2017AMS} datasets. As shown in \ref{tab:depth}, our method achieves competitive results despite having been trained on a much smaller and solely indoor dataset.

\begin{figure}[h]
    \centering
    \includegraphics[width=\linewidth]{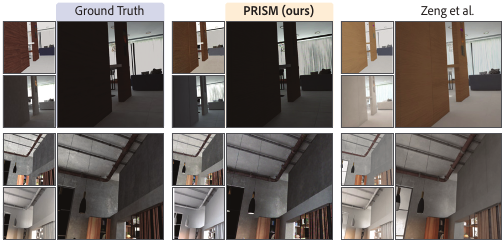}
    \caption{
        Visual comparison of RGB reconstruction from predicted albedo and irradiance for the white balance alignment. We compare \methodname model against \citet{Zeng2024RGBXID} and ground truth reconstructions. 
    }
    \label{fig:alignment}
    \vspace{-1em}
\end{figure}

Furthermore, a common issue when processing individual intrinsic maps separately is the alignment across different modalities, e.g., the white balance ambiguity between albedo and irradiance. Here, we report reconstruction error of \methodname, \methodname single channel variant where we use \methodname-albedo and \methodname-irradiance to estimate albedo and irradiance separately, as well as the strongest baseline~\citet{Zeng2024RGBXID}. Specifically, given an RGB image and the corresponding maps, we reconstruct the diffuse appearance of the target image simply by multiplying albedo and irradiance maps. We then compare the reconstructed image of the two approaches to the ground truth. The brightness of the reconstructed image will be shifted if the two channels are not aligned in terms of white balance. We show the metrics of the two approaches on the HyperSim test set in Table~\ref{tab:alignment}, as well as some visual examples in Figure~\ref{fig:alignment}. Our method generates reconstructions closer to the ground truth which we attribute to the fact that jointly generating multiple modalities result in intrinsic channels that are better aligned. We also note that \methodname single channel leads to degradation in performance compared to our final \methodname model, which shows the advantage of joint generation over single channel estimation.

\begin{table}[h]
\caption{We report the reconstruction error for \methodname, \methodname single channel variant and \citet{Zeng2024RGBXID}. The reconstruction is conducted by multiplying the predicted albedo and irradiance. Our joint inference strategy addresses the white balance ambiguity between albedo and irradiance, leading to improved quantitative and qualitative performance.}
\vspace{-1em}
\label{tab:alignment}
\begin{center}
\begin{tabular}{l|c|c|c}
\toprule
\textbf{Input} & \textbf{RMSE $\downarrow$} & \textbf{PSNR $\uparrow$} & \textbf{LPIPS $\downarrow$} \\
\midrule
\methodname & \textbf{0.0849} & \textbf{22.38} & \textbf{0.15} \\
\methodname single channel & 0.1299 & 19.94 & 0.16 \\
\citet{Zeng2024RGBXID} & 0.1437 & 16.55 & 0.16 \\
\bottomrule
\end{tabular}
\end{center}
\end{table}

\vspace{-1em}

We further evaluate our method on the out-of-domain synthetic ARAP dataset \cite{BKPB17} following the same experimental setup as CID \cite{Careaga2024ColorfulDI}  and report the results in Table \ref{tab:arap}. As shown in Table \ref{tab:arap}, our method is on-par with the SOTA method CID and better than previous diffusion-based methods.

\begin{table}[h]
\caption{Zero-shot albedo evaluation on the synthetic ARAP Dataset \cite{BKPB17}. Our proposed method estimates is on-par with SOTA CID
method, and outperforms it in terms of SSIM.}
\vspace{-1em}
\label{tab:arap}
\begin{center}
\begin{tabular}{l|ccc} 
Method & LMSE $\downarrow$ & RMSE $\downarrow$ & SSIM $\uparrow$ \\
\hline Chromaticity & 0.038 & 0.193 & 0.710 \\
Constant Shading & 0.047 & 0.264 & 0.693 \\
\hline 
\citet{luo2023crefnet} $^*$ & 0.023 & 0.129 & 0.788 \\
\citet{Kocsis2024IntrinsicID} & 0.030 & 0.160 & 0.738 \\
\citet{InteriorVerse} & 0.029 & 0.184 & 0.729 \\
\citet{OrdinalShading} & 0.035 & 0.162 & 0.751 \\
\citet{Careaga2024ColorfulDI} & $\mathbf{0.021}$ & $\mathbf{0.149}$ & 0.796 \\
\methodname  & 0.022 & $\mathbf{0.149}$ & $\mathbf{0.798}$
\end{tabular}
\end{center}
\end{table}

In addition to evaluations on synthetic test sets, we evaluate our method on two real datasets with respect to albedo estimation quality: IIW \cite{Bell2014IntrinsicII} and MAW \cite{Wu2023MeasuredAI} datasets. We first evaluate our method on the Intrinsic Images in-the-Wild (IIW) dataset, which provides pairwise human annotations of albedo brightness between sparsely sampled pixels, and present the results in Table \ref{tab:iiw}. Although our method is outperformed by \citet{Careaga2024ColorfulDI} by a close margin, we note that the transparent surfaces in the IIW dataset are marked as white, while some of our datasets treat them as black, which we think leads to some discrepancy. For the MAW dataset, which consists of $\sim$850 indoor images and measured albedo within specific masked regions in the image, we report the intensity and chromaticity of albedo estimates in Table \ref{tab:maw}.

\begin{table}
\caption{We report the WHDR error of albedo estimates on the IIW dataset for \methodname and baselines.}
\vspace{-1em}
\label{tab:iiw}
\begin{center}
\begin{tabular}{lcc}
\hline
Method & WHDR 10\%$\downarrow$ & WHDR 20\%$\downarrow$ \\
\hline
\citet{InteriorVerse}  & 34.7             & 24.1             \\
\citet{OrdinalShading}  & 24.8             & 19.2             \\
\citet{luo2023crefnet}          & \textbf{12.8}    & \textbf{10.8}    \\
\citet{Luo2024IntrinsicDiffusionJI}      & 17.9             & 13.3             \\
\citet{Kocsis2024IntrinsicID}   & 26.1             & 20.7             \\
\citet{Zeng2024RGBXID}     & 23.6      & 21.1             \\
\citet{Careaga2024ColorfulDI} & 16.8    & 15.6             \\
\methodname             & 17.2             & 15.9             \\
\hline
\end{tabular}
\end{center}
\end{table}

\begin{table}
\caption{We report the intensity and chromaticity of albedo estimates on the MAW dataset for \methodname and baselines.}
\vspace{-1em}
\label{tab:maw}
\begin{center}
\begin{tabular}{lcc}
\hline
Method & Intensity (x100)$\downarrow$ & Chromaticity$\downarrow$ \\
\hline
\citet{InteriorVerse}    & 1.44            & 4.94             \\
\citet{OrdinalShading}  & 0.57             & 6.56             \\
\citet{Kocsis2024IntrinsicID}    & 1.13       & 5.35    \\
\citet{Zeng2024RGBXID}     & 0.82      & 3.96             \\
Chen et al. [2024]      & 0.98             & 4.12             \\
\citet{Careaga2024ColorfulDI}  & \textbf{0.54} & \textbf{3.37}    \\
\methodname             & 0.71             & 3.92             \\
\hline
\end{tabular}
\end{center}
\end{table}

\begin{table}
\caption{ Zero-shot relative depth estimation. Better: AbsRel $\downarrow, \delta_1 \uparrow$. Our method achieves depth estimation results comparable to current SOTA methods despite having been trained on a fraction of the data. }
\label{tab:depth}
\resizebox{\linewidth}{!}{%
\begin{tabular}{|c|c|c|c|c|}
\hline \multirow[t]{2}{*}{Method} & \multicolumn{2}{|l|}{NYU-v2} & \multicolumn{2}{|l|}{ETH3D } \\
\hline & AbsRel $\downarrow$ & $\delta_1$ $\uparrow$ & AbsRel $\downarrow$ & $\delta_1$ $\uparrow$ \\
\hline DiverseDepth ~\cite{yin2021virtual}  & 0.117 & 0.875 & 0.228 & 0.694\\
\hline MiDaS ~\cite{Ranftl2019TowardsRM} & 0.111 & 0.885 & 0.184 & 0.752 \\
\hline LeReS ~\cite{Yin2020LearningTR} & 0.090 & 0.916 & 0.171 & 0.777 \\
\hline Omnidata ~\cite{eftekhar2021omnidata} & 0.074 & 0.945 & 0.166 & 0.778 \\
\hline DPT ~\cite{Ranftl2021} & 0.098 & 0.903 & 0.078 & 0.946 \\
\hline HDN ~\cite{zhang2022hierarchical} & 0.069 & 0.948 & 0.121 & 0.833 \\
\hline Depth Anything V1 ~\cite{depth_anything_v1} & 0.043 & 0.981 & 0.127 & 0.882 \\
\hline Depth Anything V2 ~\cite{depth_anything_v2} & 0.045 & 0.979 & 0.131 & 0.865 \\
\hline Marigold ~\cite{ke2023repurposing} & 0.055 & 0.964 & 0.065 & 0.960 \\
\hline \methodname & 0.061 & 0.922 & 0.142 & 0.836 \\
\hline
\end{tabular}
}
\end{table}

\textbf{Joint image generation.}
We evaluate the text-to-image generation performance of \methodname against our base text-to-image model using FID and text/image CLIP similarity scores in Table \ref{tab:single-vs-joint}. To this end, we generate $5000$ samples using the same text prompts both from our base model and \methodname. We use the MJHQ-30k \cite{li2024playground} dataset as the reference real images. We curate the testing text prompts from the MJHQ-30k and the testing split of the Hypersim dataset. We notice that \methodname exhibits a slight degradation in performance which can be attributed to the fact that it has been mostly trained on synthetic indoor images. Hence, we also use the interior subset of MJHQ-30k-interior that consists of mostly indoor scenes as the real reference set for computing FID. In this case \methodname performs slightly better than the base model. Overall, we conclude that the unified generation strategy does not degrade the image generation capabilities of the base model.

\textbf{Conditional image generation.}
Finally, we evaluate the conditional image generation performance of our model against commonly used conditioning baselines in Table \ref{tab:cond-gen}. In addition to popular approaches like ControlNet and T2I-Adapter, we also compare against recent methods, ControlVAR \cite{Li2024ControlVAREC} and CAR \cite{Yao2024CARCA} that also use a diffusion transformer architecture. All the baselines are trained on the same ImageNet training dataset released with ControlVAR, which is pseudo-labeled with depth and surface normals predictions. At test time, we use the provided validation split. While baselines utilize the pseuo-ground truth depth and normals provided by the dataset for conditioning, for our method we first perform intrinsic decomposition on the RGB images and use the predicted depth and normal estimates in a second pass of conditional generation. We note that our model is trained with the synthetic datasets and is tasked to generalize to the ImageNet dataset that is unseen during training in the context of this experiment. As shown in Table \ref{tab:cond-gen}, our model is on-par or superior to all the baselines in terms of the FID and IS metrics. We argue that our model's joint and aligned image generation capabilities greatly benefit controllable generation use cases.

\begin{table}[htbp] 
    \centering
    \caption{We evaluate the text-to-image generation performance of \methodname against our base text-to-image mode in terms of image quality assessment.}
    \begin{tabular}{lccc}
        \toprule
        Method & Ref. Dataset & FID $\downarrow$ & CLIP Score $\uparrow$ \\
        \midrule
        \methodname-Base  & MJHQ-30k & 19.39 & 22.80 ± 4.01 \\
        \methodname  & MJHQ-30k & 23.98 & 22.30 ± 4.79 \\
        \methodname-Base  & MJHQ-interior & 19.01 & 22.95 ± 4.02 \\
        \methodname  & MJHQ-interior & 14.98 & 23.00 ± 4.65 \\
        \methodname-Base  & HyperSim & 28.33 & 22.70 ± 4.05 \\
        \bottomrule
    \end{tabular}
    \label{tab:single-vs-joint}
\end{table}

\begin{table}[t!]
\caption{Comparison with existing controllable generation approaches on ImageNet. Our \methodname surpasses baselines with respect to FID scores despite not having seen ImageNet images during training.
}
\label{tab:cond-gen}
\begin{center}
\begin{tabular}{l|cc|cc}
\toprule
\multirow{2}{*}{Methods} & \multicolumn{2}{c|}{Depth Map} & \multicolumn{2}{c}{Normal Map} \\
& FID $\downarrow$ & IS $\uparrow$ & FID $\downarrow$ & IS $\uparrow$ \\
\midrule
T2I-Adapter & 9.9 & 133.6 & 9.5 & 142.8 \\
ControlNet & 9.2 & 150.3 & 8.9 & 155.3 \\
ControlVAR  & 6.5 & 178.5 & 6.8 & \textbf{187.7} \\
CAR  & 6.9 & \textbf{178.6} & 6.6 & 175.9 \\
\textbf{\methodname}  & \textbf{5.9} & 159.5 & \textbf{5.3} & 165.1 \\
\bottomrule
\end{tabular}
\end{center}
\end{table}

\subsection{Applications} \label{sec:applications}
\begin{figure}[h]
\includegraphics[width=\linewidth]{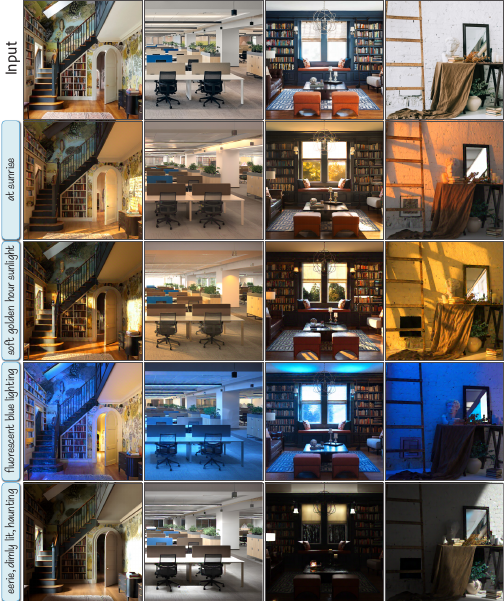}
\caption{\textbf{Relighting with text prompt.} Starting from an input RGB image, intrinsic layers are predicted using \methodname. We then apply \methodname with a text prompt describing a new light condition together with all predicted intrinsic layers except irradiance map. Our relit results preserve the geometric and material properties of the original scenes while achieving plausible appearance under desired lighting conditions.}
\label{fig:relighting}
\end{figure}

\methodname enables various global and local image editing applications through a two-step process, leveraging both its decomposition and generation capabilities. In Figure \ref{fig:relighting}, we demonstrate text guided relighting results. Given a source RGB image (top row), we first decompose it into its intrinsic components. Then, using all the decomposed intrinsic maps except the irradiance map, we perform conditional generation using a text prompt describing a new, desired light condition. As shown in the figure, our method generates re-lit images that consistently reflect the new lighting conditions  including soft shadows, specular highlights, and indirect illumination. The relit images preserve the geometric and material properties of the original scenes while generating plausible editing results.

\begin{figure}[h]
\includegraphics[width=\linewidth]{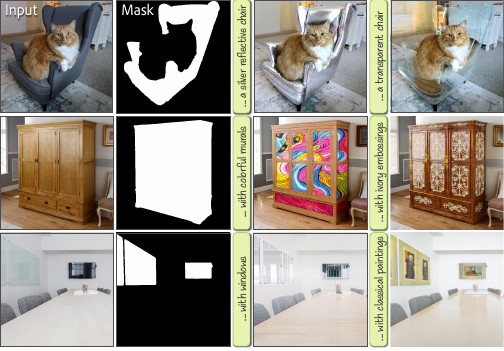}
\caption{\textbf{Material editing.} \methodname allows text conditioned material editing for a masked region of the input image. While keeping the predicted depth and irradiance fixed, albedo and normal channels are only conditioned outside the mask to allow the changes in the selected region conditioned by text prompt. Our results preserve the identity outside the mask while updating the appearance inside the mask with consistent lighting.   }
\label{fig:mat-editing}
\end{figure}

In Figure~\ref{fig:mat-editing}, we demonstrate local appearance editing results. We decompose a given input image into its intrinsic maps. We then selectively mask albedo and normal maps with random gaussian noise in the corresponding latent embedding. Using the masked maps as well as the unmasked depth and irradiance maps as a condition, we generate new images using an edited text prompt describing the desired appearance properties. The generated results preserve the identity of the scene in the unmasked regions as well as generating new looks for the masked objects while keeping the lighting consistent.

\begin{figure}[h]
\includegraphics[width=\linewidth]{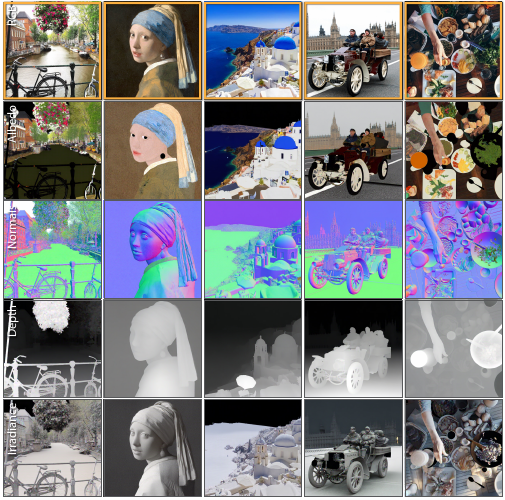}
\caption{\textbf{Out of Domain Examples.} Our fine-tune preserves the generalization of the base model. Here we show \methodname's intrinsic decomposition on out-of-domain cases such as outdoor, portrait, landscape, and top-down photography.}
\label{fig:out-of-domain}
\end{figure}

\subsection{Qualitative Results}
\begin{figure*}[h]
    \centering
    \includegraphics[width=\linewidth]{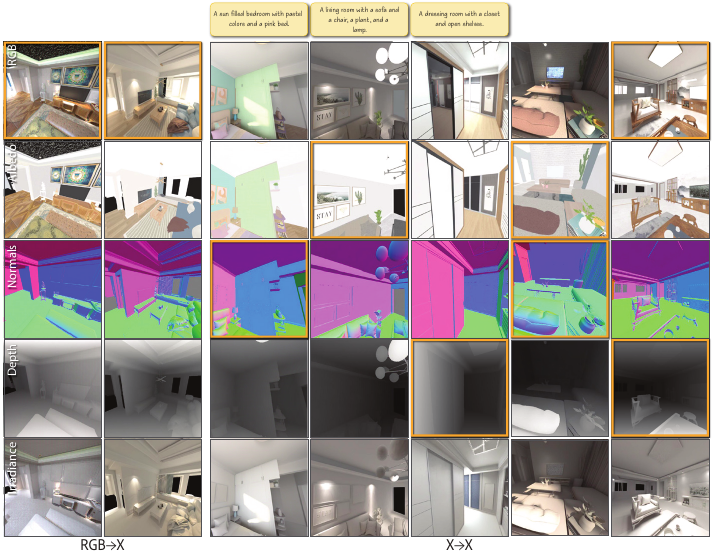}
    \caption{Sample RGB, text+X, X+RGB conditioned results generated with \methodname on the InteriorVerse test set. Condition channels are highlighted in \fcolorbox{condition}{white}{orange}.}
    \label{fig:intverse-gen-samples}
\end{figure*}

\begin{figure*}[h]
    \centering
    \includegraphics[width=\linewidth]{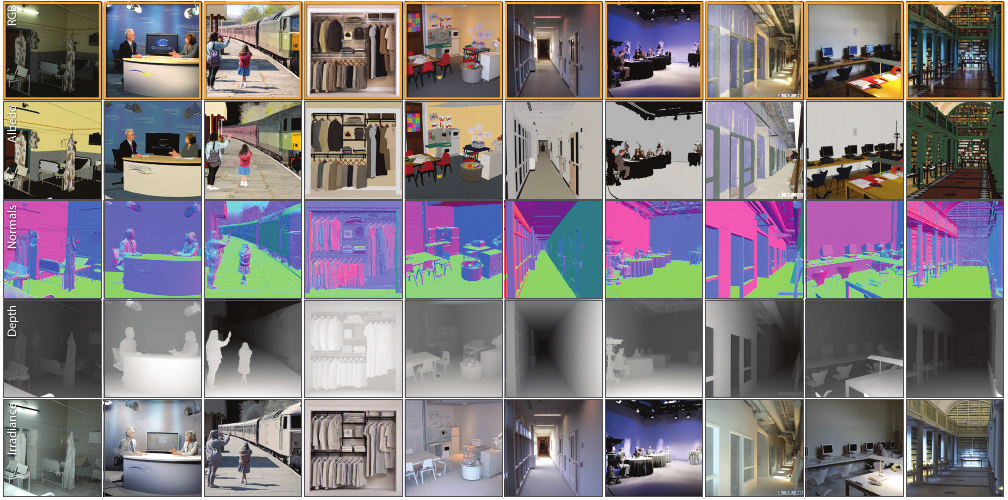}
    \caption{Additional out of domain RGB conditioned generation samples on various indoor and outdoor images. Condition channels are highlighted in \fcolorbox{condition}{white}{orange}.}
    \label{fig:out-of-domain2}
\end{figure*}

\begin{figure*}[h]
    \centering
    \includegraphics[width=\linewidth]{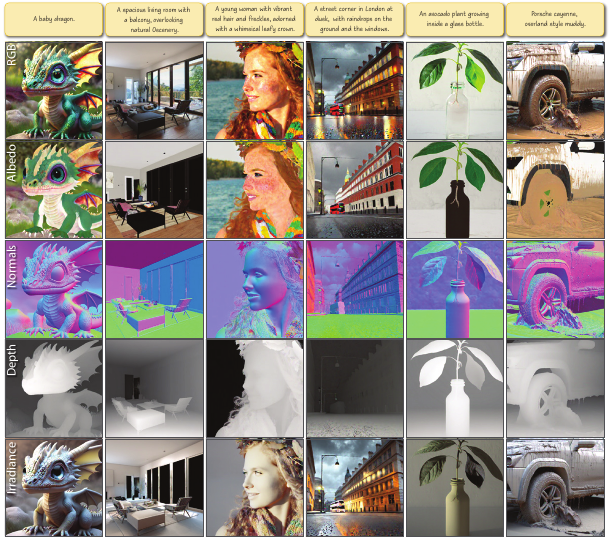}
    \caption{Text-to-RGB-X samples generated with \methodname. We use MJHQ-30k \cite{li2024playground}} text prompts from various categories to generate diverse samples.
    \label{fig:text-to-rgbx}
\end{figure*}

In Figure~\ref{fig:intverse-gen-samples}, we show more visual results from \methodname. For conditioning, we use samples from the test set of the Interior Verse dataset. We further show intrinsic image decomposition samples on real images in Figure~\ref{fig:out-of-domain} and Figure~\ref{fig:out-of-domain2}, and text-to-RGBX samples in Figure~\ref{fig:text-to-rgbx}.

\section{Conclusion}
\methodname presents a significant advancement in intrinsic image decomposition and conditional image generation, offering a unified framework that addresses key limitations of previous approaches. By generating intrinsic maps alongside RGB images within a single model, \methodname improves consistency across modalities and supports flexible, context-driven image editing tasks, including global and local adjustments. One limitation of \methodname is that it is mostly trained on interior scenes due to the availability of data with ground truth annotations. It is a promising future direction to use boost-strapping strategies where we use \methodname to annotate additional data samples for iterative training. Another future work is to explore additional modalities (e.g., segmentation masks) that are synergistic with intrinsic decomposition and can aid image editing. Finally, while we simply use updated text descriptions in a standard diffusion pass to denote desired edits, exploring how targeted generative editing paradigms such as Prompt-to-Prompt~\cite{Hertz2022PrompttoPromptIE} and InstructPix2Pix~\cite{Brooks2022InstructPix2PixLT} can be built on top of \methodname is an interesting avenue.

\bibliographystyle{ACM-Reference-Format}
\bibliography{main}

\end{document}